\documentstyle[aps,prb,twocolumn]{revtex}

\begin{document}

  \title{Model for a Josephson junction array coupled to a resonant cavity}
  \author{J.~Kent~Harbaugh and D.~Stroud}
  \address{Department of Physics, 174 W. 18th Ave., Ohio State University,
Columbus, OH 43210}
  \date{\today}
  \maketitle

\begin{abstract}
	We describe a simple Hamiltonian for an underdamped Josephson
array coupled to a single photon mode in a resonant cavity.  Using a
Hartree-like mean-field theory, we show that, for any given strength of
coupling between the photon field and the Josephson junctions, there is a
transition from incoherence to coherence as a function of $N$, the number
of Josephson junctions in the array.  Above that value of $N$, the energy
in the photon field is proportional to $N^2$, suggestive of coherent
emission.  These features remain even when the junction parameters have
some random variation from junction to junction, as expected in a real
array.  Both of these features agree with recent experiments by Barbara
{\it et al.} 
  \end{abstract}



\newcommand{\ba}{\begin{eqnarray}}
\newcommand{\ea}{\end{eqnarray}}
\newcommand{\pht}{{\text{phot}}}
\newcommand{\phs}{{\text{phase}}}
\newcommand{\ix}{{\text{int}}}
\newcommand{\lamtil}{{\tilde{\lambda}}}
\newcommand{\nbar}{{\bar{n}}}
\newcommand{\tot}{{\text{tot}}}
\newcommand{\dc}{{\text{dc}}}
\newcommand{\ac}{{\text{ac}}}
\newcommand{\fld}{{\text{field}}}
\newcommand{\der}{{\rm d}}
\newcommand{\nn}{\nonumber}

\section{Introduction} \label{sec:intro}

	Researchers have long sought to cause Josephson junction arrays to
radiate coherently.\cite{booi,jain} To achieve this goal, a standard
approach is to inject a d.c.\ current into an overdamped array.  If this
current is sufficiently large, it generates an a.c.\ voltage $V_\ac$
across the junctions, of frequency $\omega_J = 2eV_\dc/\hbar$, where
$V_\dc$ is the time-averaged voltage across the junction.\cite{josephson}
Each junction then radiates (typically at microwave frequencies).  If the
junctions are coherently phase-locked, the radiated power $P \propto N^2$,
where $N$ is the number of phase-locked junctions.  This $N^2$
proportionality is a hallmark of phase coherence.  But many difficulties
inhibit phase coherence in practice.  For example, the junctions always
have a disorder-induced spread in critical currents, which produces a
distribution of Josephson frequencies and makes phase locking
difficult.\cite{wies1,wies2,strog,hadley} Furthermore, in
small-capacitance (and underdamped) Josephson junctions, quantum phase
fluctuations inhibit phase
locking.\cite{under1,under2,under3,under4,under5,under6} Thus, until
recently, the most efficient coherent emission was found in
two-dimensional arrays of \emph{overdamped} Josephson junctions, where
quantum fluctuations are minimal. 

	Recently, Barbara {\it et al.}\ have reported a remarkable degree
of coherent emission in arrays of \emph{underdamped}
junctions.\cite{barbara,cawthorne} Their arrays were placed in a microwave
cavity, so as to couple each junction to a resonant mode of the cavity. 
If the mode has a suitable frequency and is coupled strongly enough to the
junction, it can be excited by a Josephson current through the junction. 
The power in this mode then feeds back into the other junctions, causing
the array to phase-lock and inducing a total power $P \propto N^2$.  For a
given coupling, Barbara {\it et al.}\ found that there is a threshold
number of junctions $N_c$ below which no emission was observed.  The
coupling, and hence $N_c$, could be varied by moving the array relative to
the cavity walls. 

	Barbara {\it et al.}\ interpreted their results by analogy with
the Jaynes-Cummings model\cite{jaynes,shore} of two-level atoms
interacting with a radiation field in a single-mode resonant cavity.  In
this case, each Josephson junction acts as a two-level atom; the coupling
between the ``atoms'' is provided by the induced radiation field.  A
dynamical calculation based on a model similar to that of Jaynes and
Cummings has been carried out by Bonifacio and
collaborators\cite{bonifacio1,bonifacio2} for Josephson junction arrays in
a cavity.  Their model does produce spontaneous emission into the cavity
above a threshold junction number, provided that the Heisenberg equations
are treated in a certain semi-classical limit appropriate to large numbers
of photons in the cavity. 

	In this paper, we present a simple model for the onset of phase
locking and coherent emission by an underdamped Josephson junction array
in a resonant cavity.  We also calculate the threshold for the onset of
phase coherence, using a form of mean-field theory.  Our model derives
from more conventional models of Josephson junction arrays, but treats the
interaction with the radiation field quantum-mechanically.  Within the
mean field theory, we find that for \emph{any} strength of that coupling,
there exists a threshold number of junctions $N_c$ in a linear array above
which the array is coherent.  Above that threshold, the energy in the
photon field is \emph{quadratic} in the number of junctions, as found
experimentally.\cite{barbara} The model is easily generalized to
two-dimensional arrays.  Furthermore, as we show, the threshold condition
and $N^2$ dependence of the energy in the radiation field, are both
preserved even in the presence of the disorder which will be present in
any realistic array.  Finally, the coupling constant between junctions and
radiation field can, in principle, be calculated explicitly, given the
geometry of the array and the resonant cavity. 

	The remainder of this paper is organized as follows.  In
Sec.~\ref{sec:model}, we describe our model and approximations.  Our
numerical and analytical results are presented in Sec.~\ref{sec:results}. 
Section~\ref{sec:disc} presents a brief discussion and suggestions for
future work. 

\section{Model} \label{sec:model}

\subsection{Hamiltonian}

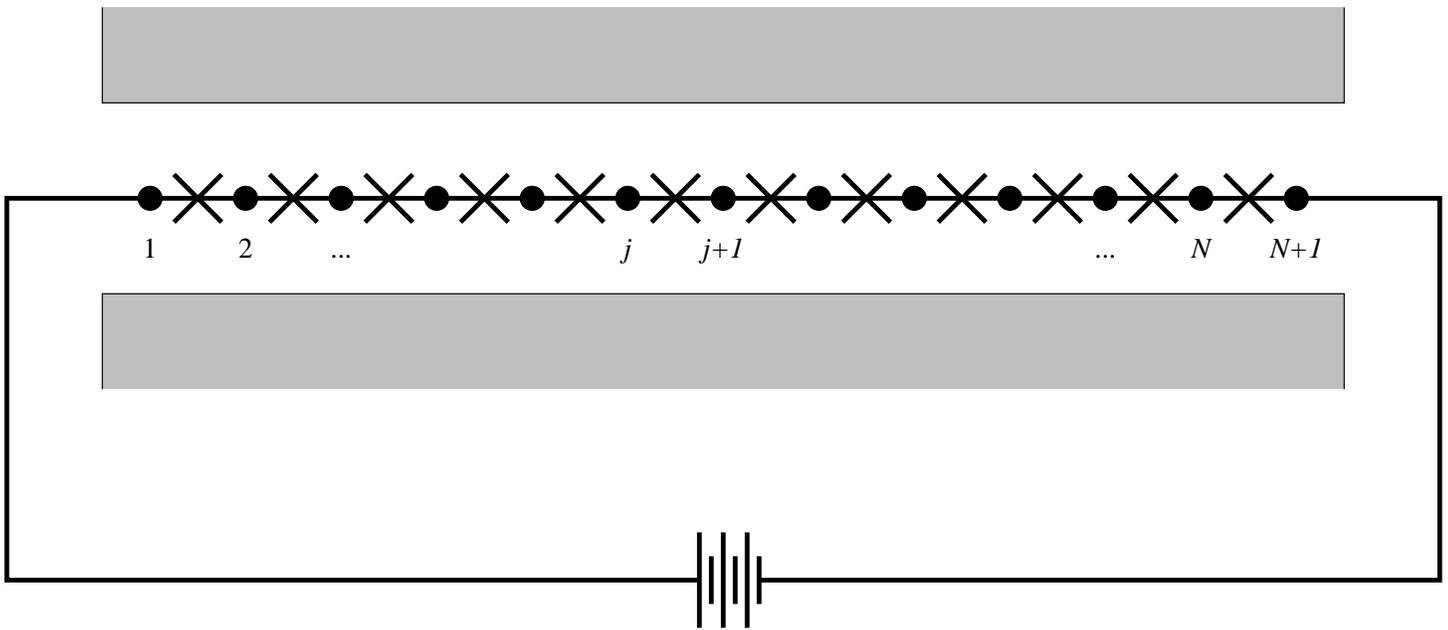
\begin{figure}
	\caption{Schematic of the geometry used in our calculations,
consisting of an underdamped array of Josephson junctions coupled to a
resonant cavity, and subjected to an applied voltage $\Phi$.  The array
consists of $N+1$ grains, represented by the dots, coupled together by
Josephson junctions, represented by the crosses.  Nearest-neighbor grains
$j$ and $j+1$ are connected by a Josephson junction, and the cavity is
assumed to support a single resonant photonic mode.  For the specific
calculations carried out in this paper, we assume a one-dimensional array
as shown, and a specific form for the capacitive energy, as discussed in
the text.}
  \label{fig:geom}
  \end{figure}

	We consider a Josephson junction array containing $N$ junctions
arranged in series, placed in a resonant cavity, arranged in a geometry
shown schematically in Fig.~\ref{fig:geom}.  It is assumed that there is a
total time-averaged voltage $\Phi$ across the chain of junctions; this
boundary condition is discussed further below.  The Hamiltonian for this
array is taken as the sum of four parts: 
  \ba
  H &=& H_J + H_C + H_{\pht} + H_\ix. 
  \label{eq:ham}
  \ea
  Here $H_J = -\sum_{j=1}^NE_{Jj}\cos\phi_{j}$ is the Josephson coupling
energy, where $\phi_j$ is the gauge-invariant phase difference across the
$j$th junction, $E_{Jj} = \hbar I_{cj}/q$, the critical current of the
$j$th Josephson junction is $I_{cj}$, and $q = 2|e|$ is the magnitude of a
Cooper pair charge.  $H_C$ is the capacitive energy of the array, which we
assume can be written in the form $H_C = \sum_{j=1}^N E_{Cj} n_j^2$, where
$E_{Cj} = q^2/(2C_j)$, the capacitance of the $j$th junction is $C_j$, and
$n_j$ is the difference in the number of Cooper pairs on the two grains
connected by the $j$th junction.  The field energy may be written as
$H_{\pht} =\hbar\Omega (a^\dag a + 1/2)$, where $\Omega$ is the frequency
of the cavity resonant mode (assumed to be the only mode supported by the
cavity), and $a^\dag$ and $a$ are the usual photon creation and
annihilation operators, satisfying the commutation relations $[a, a^\dag]
= 1$; $[a, a] = [a^\dag, a^\dag] = 0$.  We assume that the number operator
$n_j$ and phase $\phi_k$ have commutation relations $[n_j, \exp(\pm
i\phi_k)] = \pm \exp(\pm i\phi_j)\delta_{jk}$, which implies that $n_j$
can be represented as $-i\partial/(\partial\phi_j)$. 

	The crucial term in the Hamiltonian for phase locking is the
interaction term $H_\ix$.  We write this in the form $H_\ix = (1/c)
\int{\bf J}\cdot {\bf A} \der^3x$, where ${\bf J}$ is the Josephson
current density, ${\bf A}$ is the vector potential corresponding to the
electric field of the cavity mode, $c$ is the speed of light, and the
integral is carried out over the cavity volume.  Since ${\bf J}$ is
comprised of the Josephson currents $I_{cj}\sin\phi_j$ passing through the
junctions, we may write this last term as $H_\ix = \sum_{j=1}^N
E_{Jj}A_j\sin\phi_j$. 
  \ba
  A_j &=& \frac{q}{\hbar c} \int_j^{j+1} {\bf A}\cdot\der{\bf s},
  \ea
  where the integral is across the $j$th junction, i.e., between the $j$th
and $(j+1)$th superconducting grain (see Fig.~\ref{fig:geom}).\cite{note1}
The phase factor $A_j$ may be expressed in terms of the creation and
annihilation operators for the photon quanta as (in esu) $A_j =
i\sqrt{\hbar c^2/(2\Omega)}(a - a^\dag )\alpha_j$, where $\alpha_j$ is a
suitable coupling constant depending on the polarization and electric
field of the cavity mode.\cite{note2} It is convenient to introduce the
notation $\hbar g_j/\sqrt{V}= E_{Jj}\alpha_j\sqrt{\hbar c^2/(2\Omega)}$,
where $V$ is the cavity volume. 

	Finally, we need to discuss suitable boundary conditions for this
linear array.  Let $\Phi_j$ denote the time-averaged voltage across the
$j$th junction.  For our assumed form of the capacitive energy, $\Phi_j
=q\langle n_j\rangle/C_j$, where $\langle\ldots\rangle$ denotes a
quantum-statistical average.  We will impose a constant-voltage boundary
condition, by requiring that $\Phi = \sum_{j=1}^N\Phi_j$ across the linear
array should take on a specified value.  Here $\Phi$ represents the total,
time-averaged voltage across the linear array.  It is most convenient to
impose the constant-voltage boundary condition by using the method of
Lagrange multipliers, adding to the Hamiltonian a term
$\mu\sum_{j=1}^N\Phi_j = \mu\sum_{j=1}^Nqn_j/C_j$, where the constant
$\mu$ will be determined later by specifying $\Phi$. 

	If we combine all these assumptions, we can finally write an
explicit expression for $H^\prime$, the operator whose ground state we
seek: 
  \ba
  H^\prime &=& H + \mu\sum_{j=1}^Nqn_j/C_j \nn\\
  &=& \hbar\Omega \left(a^\dag a + \frac{1}{2} \right) + \sum_{j=1}^N
\biggl( -E_{Jj}\cos\phi_j + E_{Cj}n_j^2 \nn\\ 
  && \mbox{} + \mu q n_j/C_j + \frac{\hbar g_j}{\sqrt{V}}i(a-a^\dag)
\sin\phi_j\biggr). 
  \label{eq:ham1}
  \ea

\subsection{Mean-Field Approximation}

	The eigenstates of $H^\prime$ are many-body wave functions,
depending on the phase variables $\phi_j$ and $n_j$, and the photon
coordinates $a$ and $a^\dag$.  We will estimate the ground state wave
function and energy using a mean-field approximation.  To define this
approximation, we express $H^\prime$ in the form
  \ba
  H^\prime &=& H_\phs + H_{\pht} + H_\ix,
  \ea
  where $H_\phs = \sum_{j=1}^N (-E_{Jj}\cos\phi_j + E_{Cj}n_j^2 + \mu
qn_j/C_j)$, and $H_\ix = i(\hbar /\sqrt{V})(a-a^\dag) \sum_{j=1}^N
g_j\sin\phi_j$.  The mean-field approximation consists of
writing\cite{note3}
  \ba
  H_\ix &\approx& i\frac{\hbar}{\sqrt{V}} \left(\langle a - a^\dag\rangle
\sum_{j=1}^N g_j\sin\phi_j \right.\nn\\ 
  &&\mbox{} + (a - a^\dag) \sum_{j=1}^N g_j\langle\sin\phi_j\rangle \nn\\
  &&\left.\mbox{} - \langle a - a^\dag\rangle \sum_{j=1}^N
g_j\langle\sin\phi_j\rangle \right). 
  \ea
  With this approximation, $H^\prime$ is decomposed into a sum of one-body
terms, each of which depends only on the photon variables or on the phase
variables of one junction, plus a constant term.  The eigenstates of
$H^\prime$, in this approximation, are of the form $\Psi(a, a^\dag,
\{\phi_j\}) = \psi_\pht(a, a^\dag)\,\prod_{j=1}^N \psi_j(\phi_j)$, where
$\psi_\pht$ and the $\psi_j$'s are one-body wave functions. 

	That part of $H^\prime$ which depends on photon variables may be
written $H_{\pht}^\prime = H_{\pht} + i(\hbar /\sqrt{V}) (a-a^\dag)
\sum_{j=1}^N g_j\langle\sin\phi_j\rangle$, where
$\langle\sin\phi_j\rangle$ denotes a quantum-mechanical expectation value
with respect to $\psi_j(\phi_j)$.  With the definition $\lambda_j =
\langle \exp(i\phi_j)\rangle$, $H_\pht^\prime$ takes the form
  \ba
  H_\pht^\prime &=& \hbar\Omega \left(a^\dag a+\frac{1}{2} \right) +
i\sum_{j=1}^N\frac{\hbar g_j\lamtil_j}{\sqrt{V}} (a-a^\dag),
  \ea
  where $\lamtil_j = (\lambda_j-\lambda_j^*) / (2i) = \langle \sin\phi_j
\rangle$.  This is the Hamiltonian of a \emph{displaced} harmonic
oscillator; its ground state energy eigenvalue $E_{\pht;0}$ is readily
found by completing the square to obtain
  \ba
  H_\pht^\prime &=& \hbar\Omega \left(b^\dag b + \frac{1}{2}\right) -
\frac{\hbar}{\Omega V} \eta^2,
  \ea
  where $b^\dag = a^\dag + i\sum_{j=1}^Ng_j\lamtil_j/(\Omega\sqrt{V})$,
and we have defined
  \ba
  \eta &=& \sum_{j=1}^Ng_j\lamtil_j. 
  \label{eq:op}
  \ea
  (Note that $b$ and $b^\dag$ have the same commutation relations as $a$
and $a^\dag$, i.\ e., $[b, b^\dag] = 1$.)

	The resulting ground state energy of $H_\pht^\prime$ is
  \ba
  E_{\pht;0} &=& \frac{1}{2}\hbar\Omega - \frac{\hbar}{\Omega V} \eta^2. 
  \ea
  Similarly in the ground state, since $\langle b^\dag \rangle = 0$,
  \ba
  \langle a^\dag \rangle &=& -i\frac{1}{\Omega\sqrt{V}} \eta. 
  \label{eq:adag}
  \ea
  Also, the total energy stored in the photon field is
  \ba
  E_\pht\ =\ \hbar\Omega\langle a^\dag a + \frac{1}{2}\rangle &=&
\frac{\hbar\Omega}{2} + \frac{1}{\Omega V} \eta^2,
  \label{eq:ephot}
  \ea
  where we use Eq.~(\ref{eq:adag}) and the fact that $\langle b^\dag b
\rangle = 0$ in the ground state. 

	The wave function $\psi_j(\phi_j)$ is an eigenstate of the
effective single-particle Schr\"{o}dinger equation $H_j\psi_j=E_j\psi_j$,
where
  \ba
  H_j &=& -E_{Jj}\cos\phi_j + \frac{1}{2C_j}(q^2n_j^2 +2\mu qn_j) \nn\\
  &&\mbox{} + \frac{\hbar g_j}{\sqrt{V}}i \sin\phi_j \langle a -
a^\dag\rangle,
  \label{eq:schrod}
  \ea
  and $\langle a - a^\dag \rangle = 2i\eta / (\Omega\sqrt{V})$.  Using
this expression and completing the square, we can write
  \ba
  H_j &=& E_{Cj}(n_j-\bar{n})^2 - E_{Cj}\nbar^2 \nonumber\\
  &&\mbox{} - \frac{2\hbar g_j}{\Omega V} \eta\sin\phi_j -
E_{Jj}\cos\phi_j,
  \ea
  where we have written $\nbar = -\mu/q$.  (Note that with our definition
of $\mu$, it does not have the dimension of energy.) Introducing the
notation
  \ba
  E_{\alpha;j}^2 &=& E_{Jj}^2 + E_{\ix;j}^2,
  \label{eq:Ea}
  \ea
  where we define $E_{\ix;j} = 2\hbar \eta g_j / (\Omega V)$ and
$\phi_{\alpha;j} = \tan^{-1} (E_{\ix;j}/E_{Jj})$, we obtain
  \ba
  H_j &=& -E_{\alpha}\cos(\phi_j-\phi_{\alpha;j}) + E_{Cj}
\left[(n_j-\nbar)^2 - \nbar^2 \right]. 
  \label{eq:mathieu}
  \ea

	The Schr\"{o}dinger equation
  \ba
  H_j\psi_j(\phi_j) &=& E_j\psi_j(\phi_j),
  \label{eq:schrod1}
  \ea
  where $H_j$ is given by expression~(\ref{eq:mathieu}), can be
transformed into Mathieu's equation by a suitable change of variables.
Specifically, if we use the representation $n_j =
-i\partial/(\partial\phi_j)$, and we also make the change of variables
$\psi_j(\phi_j) = \exp(i\nbar\phi_j)u_j(\phi_j)$, then
Eq.~(\ref{eq:schrod1}) takes the form
  \ba
  (E_j + \bar{n}^2) u_j &=& -E_{\alpha;j} \cos(\phi_j-\phi_{\alpha;j}) u_j
- E_{Cj} \frac{\partial^2 u_j}{\partial\phi_j^2}. 
  \label{eq:mathieu1}
  \ea
  Since $\phi_j$ and $\phi_j + 2\pi$ represent the same physical state,
the physically significant eigenstate $\psi_j(\phi_j)$ should satisfy
$\psi_j(\phi_j+2\pi) = \psi_j(\phi_j)$, or equivalently
  \ba
  u_j(\phi_j+2\pi) &=& \exp(-2\pi i\bar{n})u_j(\phi_j). 
  \label{eq:bc}
  \ea
  Thus, the solutions to Eq.~(\ref{eq:mathieu1}) are Mathieu functions
satisfying the boundary condition~(\ref{eq:bc}). 

	The total ground-state energy of the coupled system takes the form
  \ba
  E_\tot &=& \sum_{j=1}^NE_{j;0} + E_{\pht;0} + E_d,
  \label{eq:etotal}
  \ea
  where $E_{j;0}$ is the lowest eigenvalue of the Schr\"{o}dinger
equation~(\ref{eq:schrod1}). Note that the $E_{j;0}$'s are also functions
of the $\lamtil_j$'s, but only through the variable $\eta$. $E_d$ is a
``double-counting correction'' which compensates for the fact that the
interaction energy is included in both $E_{\pht;0}$ and the $E_{j;0}$'s;
it is given by the negative of the expectation value of the last term on
the right-hand side of Eq.~(\ref{eq:schrod}), i.e.,
  \ba
  E_d\ =\ -i\frac{\hbar}{\sqrt{V}} \langle a - a^\dag\rangle
\sum_{j=1}^Ng_j \langle\sin\phi_j\rangle &=& \frac{2\hbar\eta^2}{\Omega
V}. 
  \ea
  Hence, the total ground-state energy is
  \ba
  E_\tot(\eta) &=& \sum_{j=1}^N E_{j;0} + E_{\pht;0} + E_d \nn\\
  &=& \sum_{j=1}^NE_{j;0}(\eta) + \frac{1}{2}\hbar\Omega +
\frac{\hbar}{\Omega V}\eta^2. 
  \label{eq:etot}
  \ea
  The actual ground-state energy is found from this expression by
minimizing $E_\tot$ with respect to the variable $\eta$, holding $\mu$ (or
$\nbar$) fixed. 

\subsection{Approximate Minimization}

	We begin by considering the case $\bar{n} = 0$, for which an
approximate minimization of $E_\tot(\eta)$ can be done analytically as
follows.  First, one must evaluate the energies $E_{j;0}$, which are the
ground-state eigenvalues of $H_j\psi_j(\phi_j) = E_j\psi_j(\phi_j)$.  For
$\bar{n} = 0$, E$_{j;0}$ has the approximate value\cite{mathieu}
  \ba
  E_{j;0} &\sim& -\frac{E_{\alpha;j}^2}{2 E_{Cj}}
  \ea
  for $E_{\alpha;j} \ll E_{Cj}$, and
  \ba
  E_{j;0} &\sim& -E_{\alpha;j}
  \ea
  for $E_{\alpha;j} \gg E_{Cj}$.  A function which interpolates smoothly
between these limits is
  \ba
  E_{j;0} &\sim& E_{Cj} - \sqrt{E_{Cj}^2 + E_{\alpha;j}^2}. 
  \ea
  Substituting this expression into Eq.~(\ref{eq:etot}), we obtain
  \ba
  E_{\tot}(\eta) &=& \frac{\hbar}{\Omega V} \eta^2 + \sum_{j=1}^N \Biggl[
E_{Cj} \nn\\
  &&\mbox{} - \sqrt{E_{Cj}^2 + E_{Jj}^2 + {\left( \frac{2\hbar g_j}{\Omega
V} \right)}^2 \eta^2} \Biggr]. 
  \label{eq:etot2}
  \ea
  Setting $\der E_{\tot}/\der\eta = 0$, we obtain the condition
  \ba
  \eta &=& \eta \frac{2\hbar}{\Omega V} \sum_j \frac{g_j^2}{\sqrt{E_{Cj}^2
+ E_{Jj}^2 + [2\hbar g_j/(\Omega V)]^2 \eta^2}}. 
  \ea
  This equation always has the solution $\eta = 0$.  If
  \ba
  \frac{2\hbar}{\Omega V} \sum_{j=1}^N \frac{g_j^2}{\sqrt{E_{Cj}^2 +
E_{Jj}^2}} &>& 1,
  \label{eq:thresh}
  \ea
  then there is also a real, nonzero solution for $\eta$.  Whenever this
solution exists, it is a minimum in the energy, and the $\eta = 0$
solution is a local maximum.  Thus, Eq.~(\ref{eq:thresh}) represents a
\emph{threshold for the onset of coherence}. 

	In the opposite limit, when $E_{\alpha;j} \gg E_{Cj}$ and
$\lamtil_j \rightarrow 1$,
  \ba
  \eta &=& \sum_{j=1}^Ng_j\lamtil_j \rightarrow \sum_{j = 1}^Ng_j. 
  \ea
  If we define $\bar{g} = \sum_{j=1}^Ng_j/N$ and $\bar{\eta} = \eta/N$,
then we see that $\bar{\eta}$ rises from zero at a threshold determined by
Eq.~(\ref{eq:thresh}) and approaches unity when the parameters
$|E_{\alpha;j}|$ are sufficiently large. 

	For $\bar{n} \neq 0$, the threshold can still be approximately
found analytically. Since $H^\prime$ is periodic in $\bar{n}$ with period
unity, one need consider only $-1/2 < \bar{n} \leq 1/2$.  In this regime,
we write $H_j$ as
  \ba
  H_j &=& -E_{\alpha;j}\cos(\phi-\phi_\alpha) + E_{Cj}
[(n-\bar{n})^2-\bar{n}^2] \nn\\
  &=& -E_{\alpha;j}\cos(\phi-\phi_\alpha) + H_j^0. 
  \ea
  The coherence threshold occurs in the small-coupling regime,
$|E_{\alpha;j}| \ll E_{Cj}$.  The desired ground-state solution can be
obtained as a perturbation expansion about the solutions to the zeroth
order Schr\"{o}dinger equation, $H_j^0\psi_j^0 = E_j^0\psi_j^0$. The
(unnormalized) solutions to this equation are $\psi_j^0 = \exp(im\phi_j)$,
corresponding to eigenvalues $E_{j}^0 = E_{Cj}[(m-\bar{n})^2 -\bar{n}^2]$,
with $m$ integer. For $|\bar{n}| < 1/2$, the ground state is $m = 0$.  The
second-order perturbation correction to this energy due to the
perturbation $H_j^\prime = -E_{\alpha;j}\cos(\phi-\phi_\alpha)$ is
  \ba
  \Delta E_j &=& \sum_{m = \pm 1} \frac{{|\langle 0| H^\prime_j
|m\rangle|}^2}{E_0-E_m},
  \ea
  where $|m\rangle$ denotes the ket corresponding to $\exp(im\phi_j)$. 
After a little algebra, it is found that $\Delta E_j =
-E_{\alpha;j}^2/[(2E_{Cj})(1-4\bar{n}^2)]$. If $|E_{\alpha;j}| \gg
E_{Cj}$, then the ground state eigenvalue $E_{j;0}$ approaches
$-E_{\alpha;j}$ as in the case $\bar{n} = 0$. The generalization of the
formula~(\ref{eq:etot2}) to the case $\nbar \neq 0$ is readily shown to be
  \ba
  E_{\tot}(\eta) &=& \frac{\hbar}{\Omega V} \eta^2 + \sum_{j=1}^N \Biggl[
\tilde{E}_{Cj} \nn\\
  &&\mbox{} - \sqrt{ \tilde{E}_{Cj}^2 + E_{Jj}^2 + {\left( \frac{2\hbar
g_j}{\Omega V} \right)}^2 \eta^2} \Biggr],
  \label{eq:etotn}
  \ea
  where $\tilde{E}_{Cj} = E_{Cj}(1-4\bar{n}^2)$. To determine $\eta$ for a
given value of $\bar{n}$, and of the $E_{Cj}$'s, $E_{Jj}$'s, and $g_j$'s,
one minimizes this energy with respect to $\eta$, as described above. 

	In practice, at any value of $\bar{n}$, and for any given
distribution of the parameters $g_j$, $C_{Jj}$, and $E_{Jj}$, one can
easily evaluate the energy numerically, using the known properties of
Mathieu functions, hence obtaining both the coherence threshold and the
value of the order parameter $\eta$.  Once $\eta$ is known, the individual
values of the $\lamtil_j$'s can obtained by numerically solving the
Schr\"{o}dinger equation (\ref{eq:schrod1}), using the Hamiltonian
(\ref{eq:schrod}) for the ground-state eigenvalue.  Finally, the
constant-voltage condition can be imposed by choosing $\mu$ so that
$\sum_{j=1}^Nq\langle n_j\rangle/C_j$ equals the time-averaged voltage
across the array. 

\section{Results} \label{sec:results}

	Although our formalism applies equally to ordered and disordered
arrays, we will present numerical results for ordered arrays only, purely
for numerical convenience.  In the ordered case, the constants $g_j$,
$E_{Cj}$, and $E_{Jj}$ are independent of $j$.  In this ordered case, we
denote the parameters $g$, $E_C = q^2/(2C)$, and $E_J$ respectively.  For
a specified value of $\nbar$, we can find the ground-state eigenvalue
$E_{j;0}$ numerically by solving Eq.~\ref{eq:schrod1}, using the
well-known properties of the Mathieu functions.  We can then minimize the
total energy $E_{\tot}$ with respect to $\eta$.  In the ordered case, as
noted, all the $\lamtil$'s are equal, and $\eta = N\lamtil$. Furthermore,
in this case, $\nbar$ is related to $\Phi$ by $\Phi = Nq\bar{n}/C$. 
Hereafter, for given values of $g$, $E_C$, $E_J$, and $\bar{n}$, we define
$\lamtil_0$ as the value of $\lamtil$ which minimizes the total energy
$E_{\tot}$. 

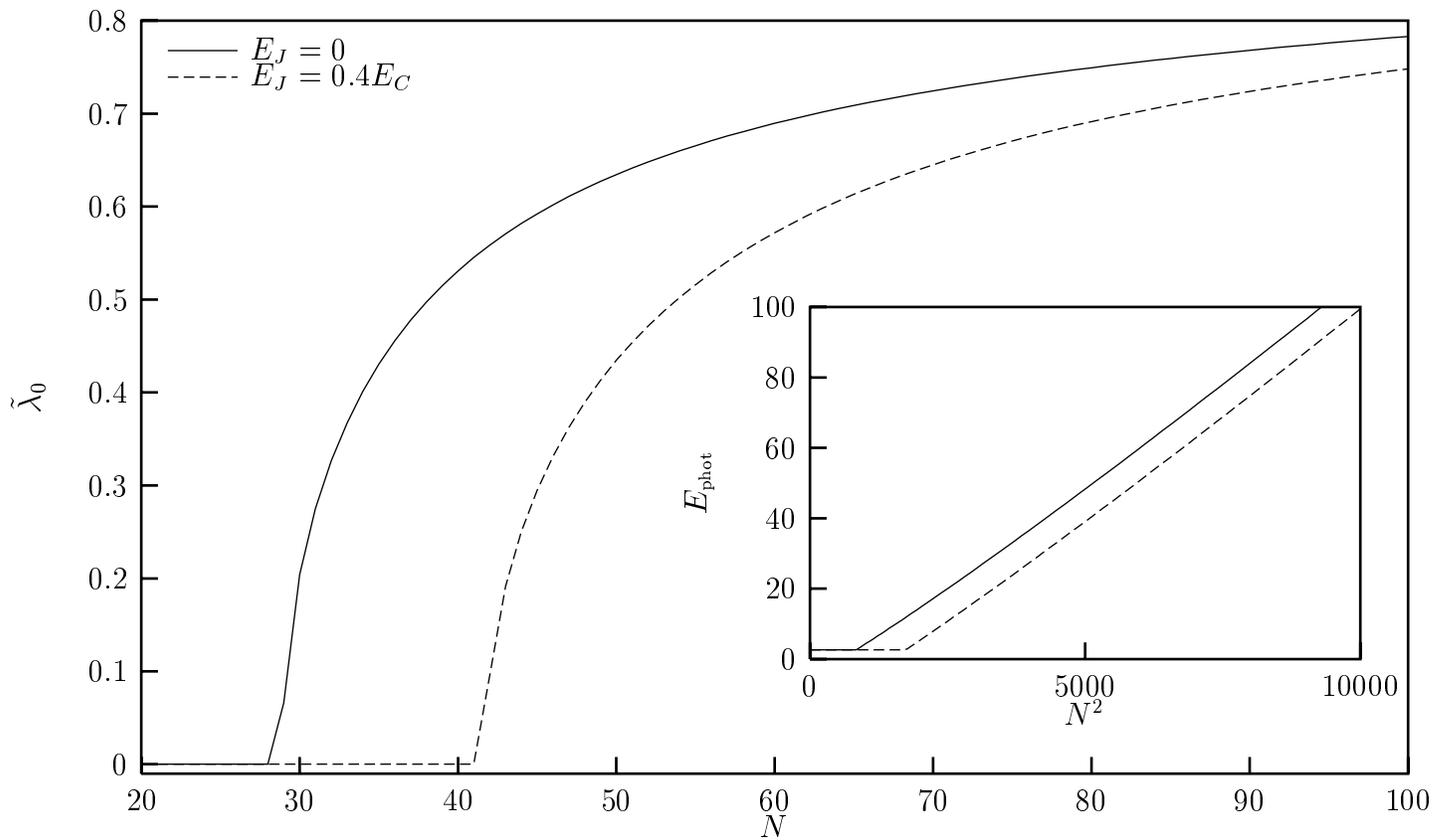
\begin{figure}
	\caption{Coherence order parameter $\lamtil_0$ which minimizes
$E_\tot(\lamtil)$ for a one-dimensional array, plotted as a function of
the number of junctions $N$, for two values of the direct Josephson
coupling energy $E_J$.  Other parameters are $\hbar g/\protect\sqrt{V}=0.3E_C$,
$\hbar\Omega/2=2.6E_C$, and $\nbar=0$.  The coupling parameter $E_{J0} =
\hbar g^2/(\Omega V)$ is given by $E_{J0}\approx 0.017E_C$.  Inset: total
energy in the photon field, $E_\pht$, plotted as a function of $N^2$, for
the same parameters and the same two values of $E_J$.}
  \label{fig:lN}
  \end{figure}

	In Fig.~\ref{fig:lN}, we plot $\lamtil_0$ for this ordered array,
as a function of $N$, assuming $\nbar = 0$.  Two curves are plotted.  The
full curve shows $\lamtil_0$ for the case $E_J = 0$, i.e., no direct
Josephson coupling.  The dashed curve in Fig.~\ref{fig:lN} shows
$\lamtil_0$ but for a finite direct Josephson coupling.  In both cases,
there is clearly a threshold array size $N_c$, below which $\lamtil_0 =
0$.  For $N > N_c$, we find $\lamtil_0 > 0$.  Since $\lamtil_0 = \langle
\sin\phi_j \rangle_0$ (that is, the expectation value of $\sin\phi_j$ in
this energy-minimizing state), the Josephson array has a net supercurrent
in this configuration.  As $N$ increases, $\lamtil_0$ approaches unity,
which corresponds to complete phase-locking.  The value of $N_c$ is larger
for finite Josephson coupling than for zero direct coupling; thus, it
appears, paradoxically, that the finite direct coupling actually impedes
the transition to coherence.  This point will be discussed further below. 

	For $E_J = 0$ and $\nbar = 0$, $N_c$ can easily be found
analytically from Eq.~(\ref{eq:thresh}).  The threshold is found to
satisfy
  \ba
  N_c &=& E_C/(2E_{J0}),
  \ea
  where $E_{J0} = \hbar g^2/(\Omega V)$.  This value agrees quite well
with our numerical results (cf.\ Fig.~\ref{fig:lN}).  Note that, for
\emph{any} nonzero value of the coupling $E_{J0}$, no matter how small,
there always exists a threshold value of $N$, above which phase coherence
becomes established. 

	If $E_J = 0$ and $\nbar = 0$, $N_c$ can still be obtained as an
implicit equation even in the disordered case, in terms of the
distribution of the $g_j$'s and $E_{Cj}$'s.  The result is readily shown
to be
  \ba
  1 &=& \frac{2\hbar}{\Omega V} \sum_{j=1}^{N_c} \frac{g_j^2}{E_{Cj}}. 
  \ea
  For a given distribution of the parameter $g_j^2/E_{Cj}$, there will
always exist a threshold value of $N$ such that this equation is
satisfied, no matter how weak the coupling constants $g_j$.  Thus, at
least in this mean-field approximation, the disorder has no qualitative
effect on the coherence transition discussed here.  In particular, the
critical number $N_c$ does not necessarily either increase or decrease
with increasing disorder; instead, $N_c$ depends on the distribution of
$g_j$, $E_{Jj}$, and $E_{Cj}$ in the array. 

	The inset to Fig.~\ref{fig:lN} shows the total energy in the
photon mode, $E_\pht = \hbar\Omega (\langle a^\dag a \rangle + 1/2)$, in
the ordered case, plotted as a function of $N$ for $\nbar = 0$.  From
Eq.~(\ref{eq:ephot}), we find that $E_\pht = \hbar\Omega/2 +
N^2\lamtil_0^2E_{J0}$ for an ordered array; this is the quantity plotted
in the inset.  As is evident from the plot, $E_\pht$ varies approximately
\emph{linearly} with $N^2$ all the way from the coherence threshold to
large values of $N$, where $\lamtil_0 \rightarrow 1$.  This $N^2$
dependence is a hallmark of phase coherence. 

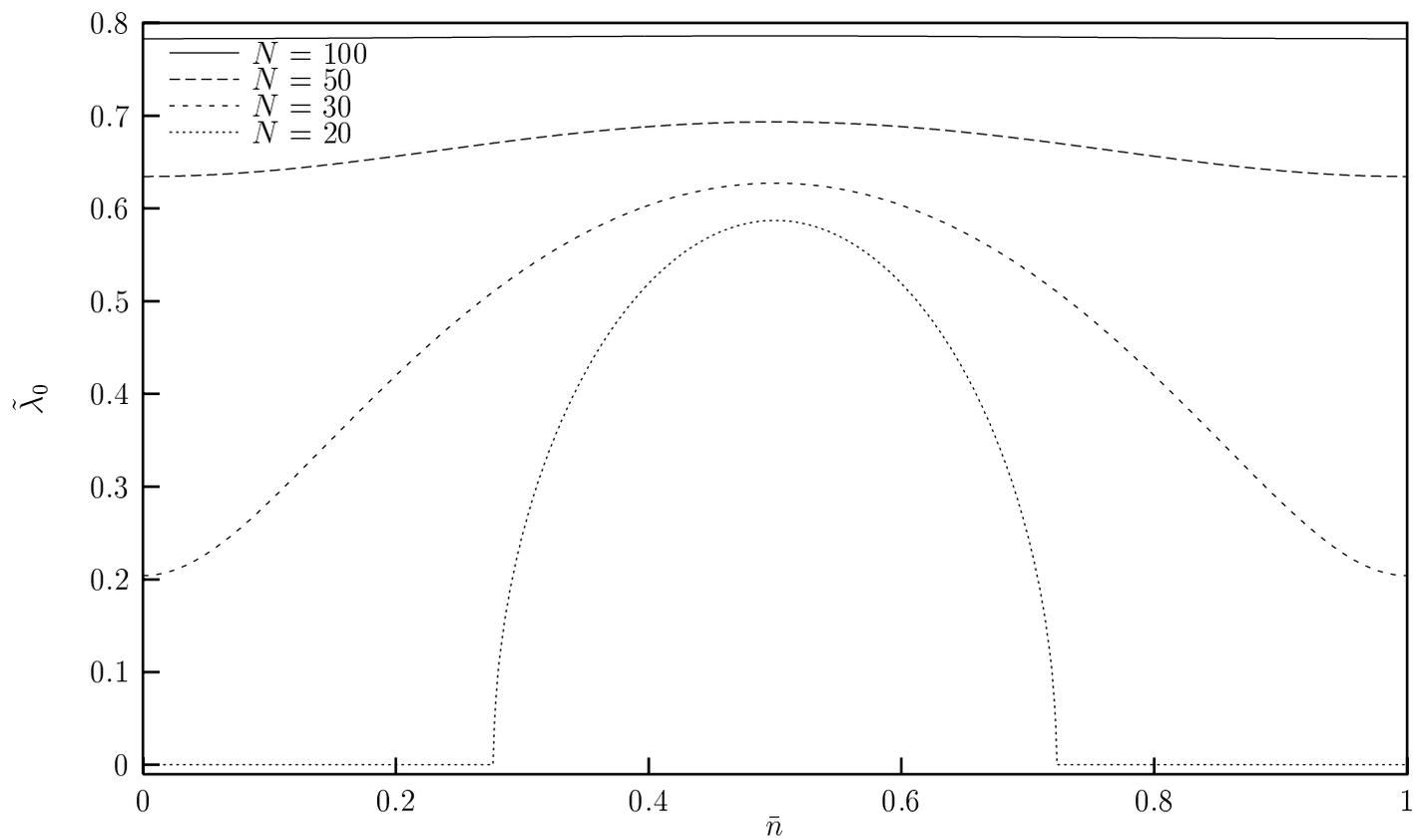
\begin{figure}
	\caption{Energy-minimizing value of the coherence order parameter
$\lamtil_0$, as a function of the parameter $\nbar=\mu/q$, for several
values of the array size $N$.  Other parameters are $\hbar
g/\protect\sqrt{V}=0.3E_C$, $\hbar\Omega/2=2.6E_C$, $E_J=0$, and $E_{J0} \approx
0.017E_C$.}
  \label{fig:ln}
  \end{figure}

	The voltage $\Phi$ across the array is determined by $\nbar$ (or
equivalently $\mu$).  In Fig.\ \ref{fig:ln}, we plot $\lamtil_0$ as a
function of $\nbar$ for several array sizes at fixed coupling constants
$E_{J0}$, $E_C$, and $E_J$ in an ordered array.  Since as already shown,
$\lamtil_0$ is periodic in $\nbar$ with a period of unity, we plot
$\lamtil_0(\nbar)$ only for a single period, $0 \leq \nbar \leq 1$.  Fig.\
\ref{fig:ln} shows that, for any given $N$ and $E_{J0}$, the calculated
$\lamtil_0$ achieves its maximum value when $\nbar$ has a half-integer
value, i.e., the array is most easily made coherent at such values of
$\nbar$. In particular, an array whose size is slightly below the
threshold value at integer values of $\nbar$ can be made to become
coherent, with a nonzero $\lamtil_0$, when $\nbar$ is increased---that is,
when a suitable voltage is applied.  On the other hand, for values of $N$
far above the threshold, $\lamtil_0$ is little affected by a change in
$\nbar$. 

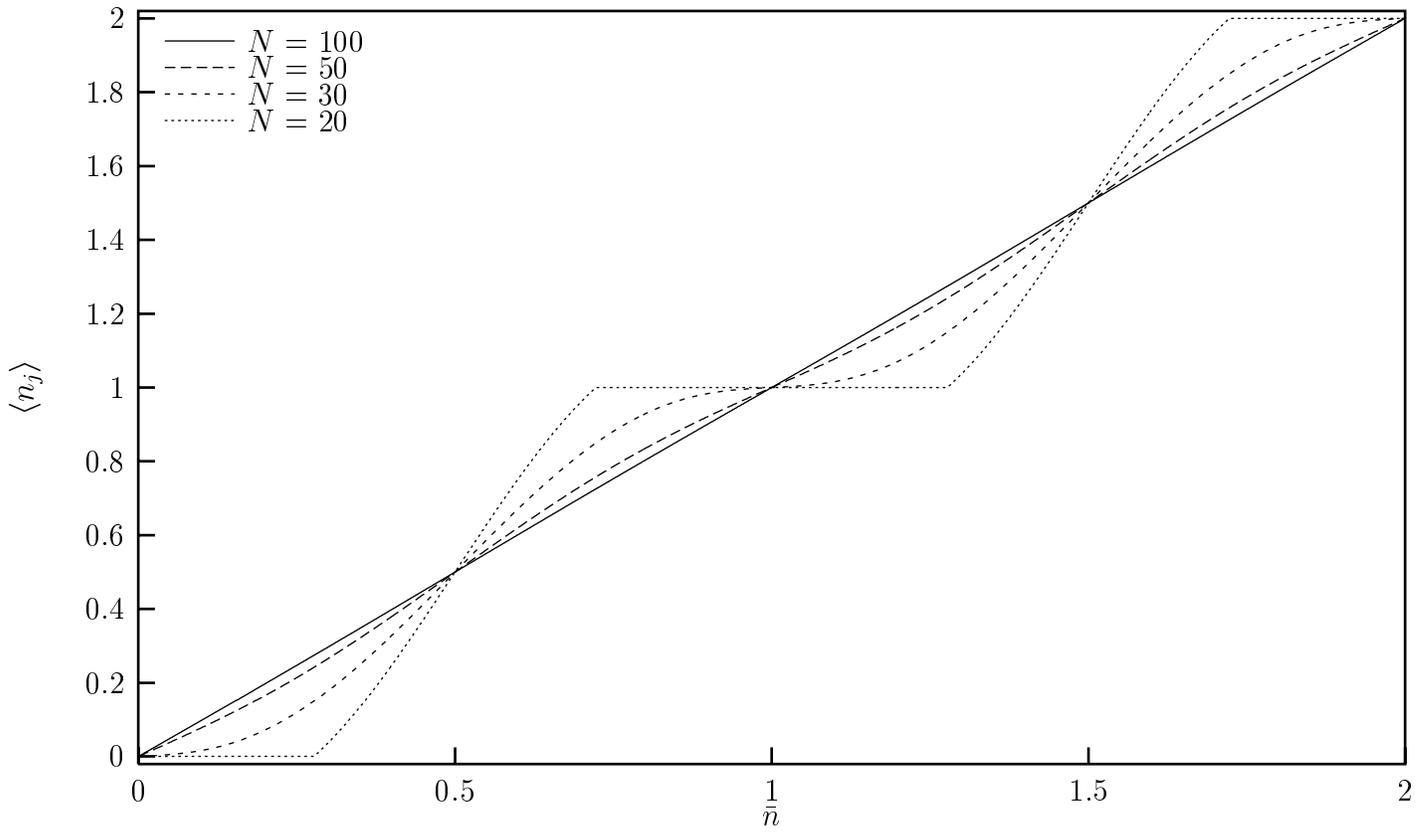
\begin{figure}
	\caption{The parameter $\langle n_j \rangle =\Phi C/(qN)$, where
$\Phi/N$ is the voltage drop across one junction, plotted as a function of
the parameter $\nbar=\mu/q$, for several values of the array size $N$. 
Other parameters are $\hbar g/\protect\sqrt{V}=0.3E_C$, $\hbar\Omega/2=2.6E_C$,
$E_J=0$, and $E_{J0}\approx 0.017E_C$.}
  \label{fig:vn}
  \end{figure}

	In Fig.~\ref{fig:vn}, we show the quantity $\langle n_j\rangle$ as
a function of $\nbar$, for several values of $N$ and fixed value of the
coupling constant ratios $E_{J0}/E_C$ and $E_J/E_C$, for a single cycle
($0 \leq \nbar \leq 1$) .  This quantity is related to the voltage drop
across one junction, in our model, by $\Phi/N = q\langle n_j\rangle/C$. 
For sufficiently large arrays, $\langle n_j\rangle \sim \nbar$ and the
voltage drop is nearly linear in $\nbar$ in this mean-field approximation. 
For arrays closer to the coherence threshold, $\langle n_j \rangle$ is a
highly nonlinear function of $\nbar$.  However, the deviation from
linearity, $\langle n_j \rangle - \nbar$, is, once again, a periodic
function of $\nbar$ with period unity.  The discontinuous jumps in $\nbar$
as a function of $\langle n_j \rangle$ represent regions of incoherence
($\lamtil_0 = 0$), whereas the regions in which $\langle n_j \rangle$ is a
smooth function of $\nbar$ are regimes of phase coherence ($\lamtil_0 \neq
0$). 

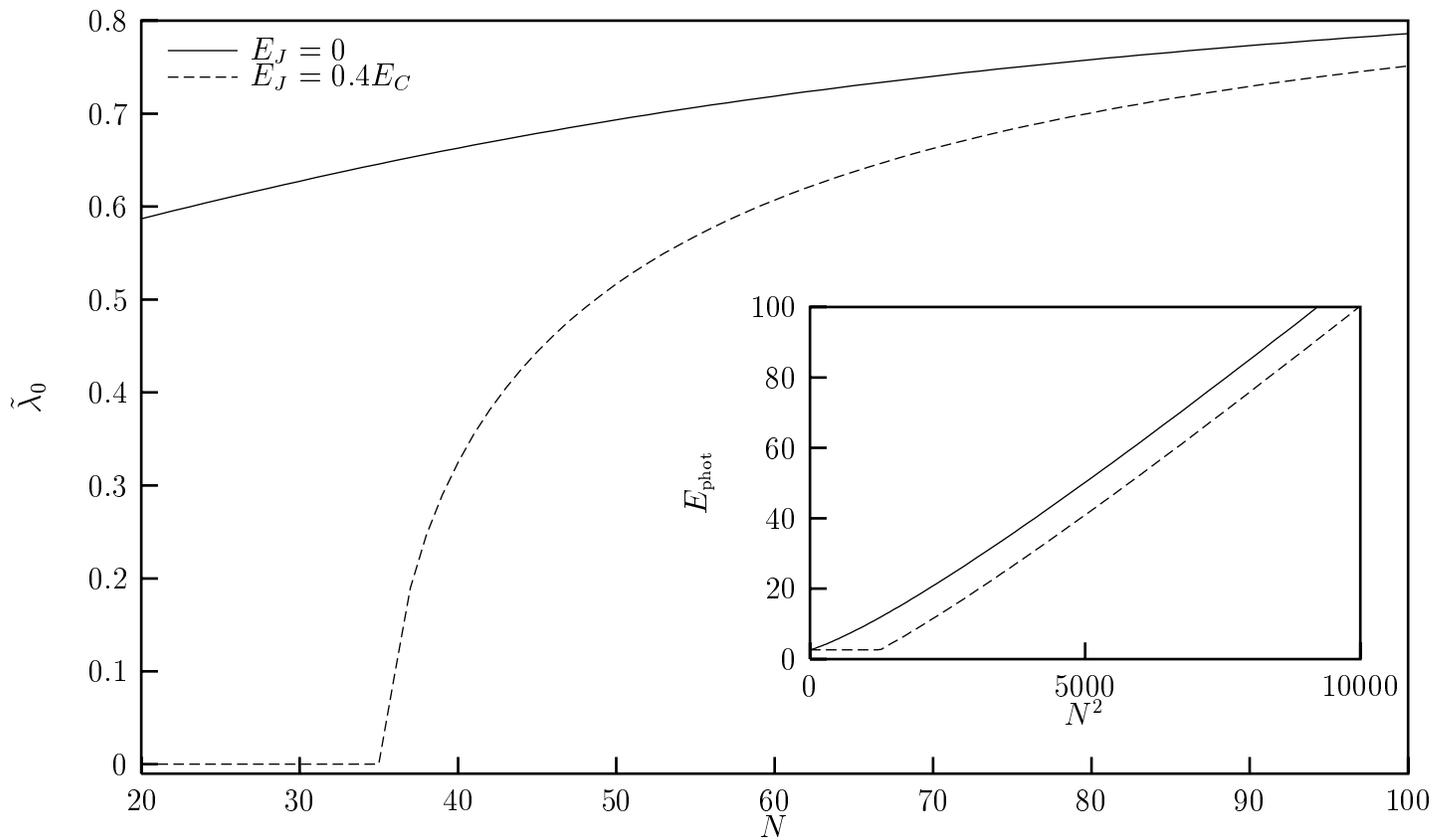
\begin{figure}
	\caption{Same as Fig.~\protect\ref{fig:lN}, but for $\bar{n} =
1/2$.}
  \label{fig:lNn}
  \end{figure}

	In Fig.~\ref{fig:lNn}, we again plot $\lamtil_0(N)$ for two fixed
ratios $E_J/E_C$, but this time for $\nbar = 1/2$.  From
Fig.~\ref{fig:ln}, we expect this choice of $\nbar$ to maximize the
tendency to phase coherence and thus to reduce the threshold array size
for the onset of phase coherence.  Indeed, in the absence of direct
Josephson coupling, this threshold is reduced to below unity (that is,
$\lamtil_0$ remains nonzero, even at $N = 1$, for our choice of $E_{J0}$). 
In fact, for this value of $\nbar$, only an infinitesimal coupling to the
resonant mode is required to induce phase coherence in this model.  Once
again (cf.\ Fig.~\ref{fig:lNn}), the addition of a finite direct Josephson
coupling actually increases the threshold number for phase coherence at
$\bar{n} = 1/2$ as it does at $\bar{n} = 0$. 

	Although we have not carried out a similar series of calculations
for a disordered, our \emph{analytical} results show that the essential
features found in the ordered case will be preserved also in a disordered
array.  Most importantly, there remains a critical junction number for
phase coherence in a disordered array, just as there does in the ordered
case.  The most important difference between the two cases is that the
individual $\lamtil_j$'s will be functions of $j$ in the disordered case. 

\section{Discussion} \label{sec:disc}

	Although the present work is only a mean-field approximation, we
expect that it will be quite accurate for large $N$.  The reason is that,
in this model, the one photonic degree of freedom is coupled to every
phase difference, and thus experiences an environment which is very close
to the mean, whatever the state of the individual junctions.  Such small
fluctuations are necessary in order for a mean-field approach to be
accurate.  In fact, a similar approach has proven very successful in work
on novel Josephson arrays in which each wire is coupled to a large number
of other wires via Josephson tunneling.\cite{vinokur,tinkham}

	It may appear surprising that a finite direct Josephson coupling
actually increases the threshold array size for coherence. But in fact
this behavior is reasonable.  If there is no direct coupling ($E_J = 0$),
the phase difference across each junction evolves independently, except
for the global coupling to the resonant photon mode.  When the array
exceeds its critical size, this coupling produces coherence.  If the same
array now has a finite $E_J$, there are two coupling terms. But these are
not simply additive, but in fact are $\pi/2$ out of phase: the direct
coupling favors $\phi_j = 0$, while the photonic one favors $\phi_j =
\pi/2$.  For a large enough array, the coupling to the photon field still
predominates and produces global phase coherence, but this occurs at a
higher threshold, at least in our model, than in the absence of direct
coupling. 

	A striking feature of our results is the very low coherence
threshold ($N = 1$) when $\nbar = 1/2$.  In fact, for any $N$ and for $E_J
= 0$, only an infinitesimal coupling to the cavity mode would be required
to induce phase coherence at $\nbar = 1/2$.  The reason for this low
threshold is that, in the absence of coupling, junction states with
$\langle n_j\rangle = 0$ and $\langle n_j \rangle = 1$ are degenerate. 
Any coupling is therefore sufficient to break the degeneracy and produce
phase coherence. A related effect has been noted previously in studies of
more conventional Josephson junction arrays in the presence of an offset
voltage.\cite{offset}

	Finally, we comment on what is not included in the present work. 
This paper really considers only the minimum energy state of the coupled
photon/junction array system under the assumption that a particular
voltage is applied across the array.  It would be of equal or greater
interest to consider the dynamical response of such an array. 
Specifically, it would be valuable to develop and solve a set of coupled
dynamical equations which incorporate both the junction and the photonic
degrees of freedom.  Such a set of equations has already been proposed by
Bonifacio under a particular set of simplifying assumptions.  A more
accurate set of equations is needed, which would include not only a
driving current, but also the damping arising from both resistive losses
in the junctions and losses due to the finite $Q$ of the cavity.  In the
absence of damping, such equations can be written down from the Heisenberg
equations of motion.  The inclusion of damping may be more difficult.  We
hope to discuss some of these effects in a future publication. 

\section*{Acknowledgments}

	One of us (DS) thanks the Aspen Center for Physics for its
hospitality while parts of this paper were being written, and acknowledges
valuable conversations with Carlos Sa de Melo and Alan Dorsey.  This work
has been supported by NSF grant DMR 97-31511 and by the Midwest
Superconductivity Consortium through DOE grant DE-FG 02-90-45427.

\begin{references}
\bibitem{note1} This form for $H_\ix$ could also have been obtained by
taking the form $-\sum_{j}E_{j}\cos(\phi_j - A_{j})$ appropriate for the
Josephson coupling energy in the presence of a vector potential, and
expanding it to first order in the $A_{j}$'s.
\bibitem{note2} The specific form of $\alpha_j$ is readily obtained from
standard expressions for ${\bf A}$ in terms of $a$ and $a^\dag$.  If we
choose a normalization such that $\int_V {|{\bf E}_a|}^2 \der^3x = 1$,
where $V$ is the volume of the cavity and ${\bf E}_a({\bf x})$ is the
electric field of the normal mode, then we find that $\alpha_j = q/(\hbar
c) \sqrt{4\pi} \int_j^{j+1} {\bf E}_a\cdot\der{\bf s}$.
\end{references}
\end{document}